\documentclass[12pt]{iopart}

\usepackage{iopams}  
\usepackage{graphicx}
\usepackage{xcolor}

\newcommand{\bq}{\begin{equation}}
\newcommand{\eq}{\end{equation}}
\newcommand{\bqn}{\begin{eqnarray}}
\newcommand{\eqn}{\end{eqnarray}}
\newcommand{\nb}{\nonumber}
\newcommand{\lb}{\label}

\begin{document}

\title{TRANSLATION IN CYLINDRICALLY SYMMETRIC VACUUM}

\author{M.-N. C\'el\'erier$^1$\thanks{marie-noelle.celerier@obspm.fr}, R. Chan$^2$\thanks{chan@on.br}, M.F.A. da Silva$^3$\thanks{mfasnic@gmail.com} and N.O. Santos$^{4,5}$\thanks{n.o.santos@qmul.ac.uk}\\
\small{$^{1}$Observatoire de Paris-Meudon, 5, Place Jules Janssen, F-92195 Meudon Cedex, France}\\
\small{$^{2}$Coordena\c{c}\~ao de Astronomia e Astrof\'{\i}sica, Observat\'orio Nacional, Rua General Jos\'e Cristino,77,}\\
\small{S\~ao Crist\'ov\~ao 20921-400, Rio de Janeiro, RJ, Brazil}\\ 
\small{$^{3}$Departamento de F\'{\i}sica Te\'orica,Instituto de F\'{\i}sica, Universidade do Estado do Rio de Janeiro,}\\
\small{Rua S\~ao Francisco Xavier 524, Maracan\~a 20550-900, Rio de Janeiro, RJ, Brazil}\\
\small{$^{4}$Sorbonne Universit\'e, UPMC Universit\'e Paris 06, LERMA, UMRS8112 CNRS,}\\
\small{Observatoire de Paris-Meudon, 5, Place Jules Janssen, F-92195 Meudon Cedex, France}\\
\small{$^{5}$School of Mathematical Sciences, Queen Mary,}\\
\small{University of London, Mile End Road, London E1 4NS, UK}}

\maketitle

\begin{abstract}

{Starting from the stationary cylindrically symmetric solution, but with the coordinates $z$ and $\phi$ interchanged, and supposing that it could describe the vacuum spacetime of a translating cylinder, we investigate its physical and geometrical properties.} {This hypothesis is not entirely new since it has already been considered in a previous paper describing a translating source}. We  show that this metric is geometrically related to the vacuum field produced by a stationary {rotating} cylindrical source{, known as Lewis solution}. However, we find new physical properties, different from those of the Lewis vacuum solution. Moreover, we show that no translating cylindrical dust solution can exist in General Relativity.

\end{abstract}

\maketitle

\section{Introduction}

In the context of general relativity, cylindrically symmetric spacetimes have aroused great interest since they allow to study a wide range of  physical systems, some of them exhibiting intrinsic symmetry related characteristics (see e.g. \cite{100} and references there in).

For instance, the difference between the Newtonian and Einsteinian gravitational viewpoints fully emerges in this context.
Even in the simple case describing the vacuum field exterior to an infinite static cylinder of matter, the Levi-Civita solution \cite{101}, this difference is obvious. In its general relativistic form, this solution contains two independent parameters \cite{102,103,104,106}, one describing the Newtonian energy per unit length of the source, and the other related to the angle defect, at variance with its Newtonian counterpart only exhibiting the first parameter. The importance of the second one emerges from its global topological meaning since it cannot be removed by scale transformations \cite{102,118,119}. It produces a gravitational analog to the Aharanov-Bohm effect which allows a (Newtonian) non observable quantity (the additional constant potential to the Newtonian potential) to become observable in the relativistic theory through angular deficit strings \cite{118, 119,105,132, 149, Muriano, LewisWeyl}.

However, the first independent parameter, understood as the Newtonian mass per unit length for small matter densities, revealed as the most elusive regarding its interpretation for higher mass densities. At these densities there are a number of obstacles and apparent contradictory properties, allowing for different possible interpretations (see a discussion in \cite{102,104,507}.

Linet \cite{160} and Tian \cite{170} presented the generalization of the Levi-Civita spacetime to include the cosmological constant $\Lambda$. It has been shown that the presence of the cosmological constant modifies drastically spacetime \cite{100,170,180,190,200,221,500}, as for instance its conformal properties.

The introduction of stationarity into a cylindrically symmetric spacetime was performed by Lewis \cite{31}, obtaining new independent parameters. Furthermore, it has been shown that rotation gives rise to two families of spacetimes, one with a flat Minkowski spacetime limit and the other without \cite{33,34,35,36}. 
Later on, Krasinski \cite{Krasinski} and Santos \cite{Santos} introduced the cosmological constant to these spacetimes.

The field equations of the time dependent vacuum spacetime with cylindrical symmetry have been obtained by Einstein and Rosen \cite{Einstein}. They describe the exterior spacetime to a collapsing cylindrical source \cite{229,217}.

All these spacetimes have been widely studied through their geometrical properties, their limits, their particle geodesics and sources. The results, some of them so weird as so far still lacking interpretations (like the relationship between the Levi-Civita, Gamma and Schwarzschild metrics \cite{Herrera1,Herrera2}), can be found in the articles here cited.

Now, the cylindrically symmetric translating source has not, to our knowledge, been studied in the literature excepted for one paper by Griffiths and Santos \cite{225} studying rotation and translation of cylinders in General Relativity. Starting from the initial hypothesis that the considered metric describes the field of cylinders in translation, it is shown there that, for an infinitely rigidly translating cylinder of perfect fluid with a regular axis, there exists a translating frame of reference relative to which the gravitational field is static. In spite of the fact that this result concurs, as its authors state, with na\"{\i}ve Newtonian intuition, it was still needed to be established in General Relativity. On the other hand, if the source exhibits a non rigid, corresponding to a non zero shear, translation, {we can ask if there is an} external field where translation can be transformed away.

In spite of the vacuum spacetime of a translating non zero shear motion source having a similar metric form to a vacuum stationary metric produced by a rotating source through interchanging its angular and axial coordinates, the geometrical and physical properties are quite different. As proved in \cite{225} shear free translating sources can be transformed, by using an appropriate frame, into a globally static field. However, a shear free rotating source cannot have its rotation transformed globally away to a static spacetime. This difference lies in the fact that rotation produces a centrifugal force that can withstand the gravitational attraction. Furthermore, for rotation, there are pressureless sources which means dust sources, while, for translation, pressure is needed to forbid the collapse of the source, as demonstrated here, in appendix A. Hence, these facts show that translation should produce some quite different results as compared to its rotation counterpart with interchanged angular and axial coordinates.

It is worth noticing that the geometrical similarity between rotation and translation in a cylindrical system reminds 
the properties of the G\"odel spacetime \cite{Godel}, in the sense that two distinct sources could imply similar geometries. There, the spacetime can be interpreted as produced by an energy momentum tensor describing either a perfect fluid or merely dust with a cosmological constant, both sources exhibiting rigid rotation. It has also been demonstrated \cite{BSM} that the G\"odel metric can be matched to an exterior one described by the stationary cylindrical Lewis vacuum spacetime with a cosmological constant \cite{Santos}. Further properties of the 
matching of these two spacetimes have been obtained by \cite{GS10}.

In the next section, we {present that one which could represent} the vacuum field solution for such a translating source. In section 3, we show, through the study of its static limit and of its geodesics, that its physical properties do not reduce to those proceeding from a mere exchange of axial and azimuthal coordinates and in section 4, we demonstrate that no translating cylindrical {dust} solution can exist. Section 5 is devoted to our conclusions.

{\section{Translation and the Lewis metric}}

{
{The most} general metric  for  cylindrically symmetric spacetimes  \cite{Bronnikov2019}, {is given by}
\bqn
\lb{metric_general}
ds^2 &=& e^{2k}dt^2 - e^{2h}dx^2 - \nb  \\
&&W^2\left[f\left(dy + w_1 dz + w_2 dx\right)^2 + f^{-1}\left(dz + w_3 dx\right)^2\right],
\eqn
where $k, \; h,\; W,\; f$, $w_1$, $w_2$ and $w_3$ are all functions of $t$ and $x$ only.
In particular, if we assume that  $w_1 = w_2 = 0$, {it is easy to see that}  (\ref{metric_general}) reduces to
\bq
\lb{metric_general_a}
ds^2 =    e^{2k}dt^2 - e^{2h}dx^2 - W^2 f d\phi^2 + W^2 f^{-1} \left( dz + w_3 dx \right)^2.
\eq
Note that in writing the above expression, we had made the replacement $(y, z) \rightarrow (\phi, z)$.
{Now, we consider a general transformation given by}
\bqn
\lb{metric_general_b}
t &=& {\bf f}(t', r), \nb \\
x &=& {\bf g}(t', r),
\eqn
where ${\bf f}(t', r)$ and ${\bf g}(t', r)$ are arbitrary functions of their indicated arguments.
If we {impose} that these new metric components, $g'_{ij}$, must obey $(\partial g'_{ij} / \partial t')=0$ in order to obtain a stationary spacetime (eliminating the undesirable cross terms and renaming $t'$ to $t$),
then it is possible to  put (\ref{metric_general}) into the form} 
\bq
ds^2=-A(\rho)dt^2+B(\rho)(d\rho^2+d\phi^2)+2k(\rho)dtdz+C(\rho)dz^2,
\lb{new-metric}
\eq
{since ${\bf f}(t, r)={\bf f}(r)$ and ${\bf g}(t, r)={\bf g}(t)$. We can note that} $A(\rho)$, $B(\rho)$, $C(\rho)$ and $k(\rho)$ can be given in terms of the Lewis parameters $a$, $b$, $c$,  and $n$ as follow
\begin{eqnarray}
A(\rho)&=& a\rho^{-n+1}-\frac{c^2}{n^2 a}\rho^{n+1}, \lb{a1} \\
B(\rho)&=&\rho^{(n^2-1)/2}, \lb{b1} \\
C(\rho)&=&-a b^2\rho^{-n+1}+\left( \frac{1}{a} \pm \frac{2b c}{n a}+\frac{b^2 c^2}{n^2 a}\right)\rho^{n+1}, \lb{c1}\\
k(\rho)&=&-a b\rho^{-n+1}+\left(\pm \frac{c}{n a}+\frac{b c^2}{n^2 a}\right)\rho^{n+1}.\lb{k1}
\end{eqnarray}

{We number the coordinates $(t,r,z,\phi)$ as $(0,1,2,3)$ and the ranges of the coordinates are $t \ge 0$,
$r \ge 0$, $-\infty < z < +\infty$ and $0 \le \phi \le 2\pi$, in order to represent cylindrical
symmetry, as { shown} in the following Section.}

{{ Although this solution is
mathematically identical to Lewis', even exhibiting the Weyl and Lewis classes, the same is not true concerning
their physical and geometrical interpretations. In this work, we investigate the possibility for metric (\ref{new-metric}),
equivalently metric (\ref{metric_general}), to describe vacuum spacetimes for translating cylinders, as supposed by Griffiths and
Santos \cite{225}. While these authors analysed the properties of a translating cylinder of matter, we study here
the behaviour of a test particle in vacuum through its geodesic motions. We use different coordinates from
those in \cite{225} in order to have a more direct analogy with the Lewis solution.}}

\section{Physical Analysis}

\subsection{{The Levi-Civita limit}}

The static Levi-Civita spacetime can be recovered for $k(\rho)=0$,  i.e., $b=c=0$. We thus denote $n=1-4\sigma$. The static version of metric (\ref{new-metric}) can now  be written
\bqn
ds^2=-a\rho^{4\sigma}dt^2+\alpha\rho^{4\sigma(2\sigma-1)}(d\rho^2+d\phi^2)+\frac{1}{a}\rho^{2(1-2\sigma)}dz^2,
\lb{metric3}
\eqn
to be compared to the Levi-Civita metric written as \cite{BSM}
\bqn
ds^2=-\rho^{4\sigma}dt^2+\rho^{4\sigma(2\sigma-1)}\left(d\rho^2+\frac{1}{a_m}dm^2\right)+\frac{1}{a_n}\rho^{2(1-2\sigma)}dn^2.
\lb{LCmetric}
\eqn
According to \cite{BSM}, if we consider $1/2<\sigma<\infty$, then we can interpret $m$ as the angular coordinate $\phi$, and $n$ (not to be mistaken for our own above relabelled integration constant $n$) as an axial coordinate $z$. This means that the parameter range $1/2<\sigma<\infty$ is equivalent to the $0<\sigma<1/2$ range with the $z$ and $\phi$ coordinates switching their nature. For the former range, the Levi-Civita metric reads
\bqn
ds^2=-\rho^{4\sigma}dt^2+\rho^{4\sigma(2\sigma-1)}\left(d\rho^2+\frac{1}{a_m}d\phi^2\right)+\frac{1}{a_n}\rho^{2(1-2\sigma)}dz^2.
\lb{LCmetric2}
\eqn

\subsection{{Geodesic equations}}

In the absence of knowledge of theoretical models for cylindrically symmetrical sources with translational motion, the geodetic motion of a test particle can be an important tool to { analyse} the effects generated by these sources in the external vacuum space. Considering metric (\ref{new-metric}), the geodesic equations are

\bqn
\ddot{t} + \left(\frac{A'C + kk'}{D^2}\dot{t} + \frac{kC' - Ck'}{D^2}\dot{z}\right)\dot{\rho} = 0,\label{geot}
\eqn

\bqn
\ddot{\rho} + \frac{A'}{2B}\dot{t}^2 - \frac{k'}{B}\dot{t}\dot{z} + \frac{B'}{2B}\dot{\rho}^2 - \frac{C'}{2B}\dot{z}^2 - \frac{B'}{2B}\dot{\phi}^2 = 0,\label{georho}
\eqn

\bqn
\ddot{z} + \left(\frac{Ak' - A'k}{D^2}\dot{t} + \frac{AC' + kk'}{D^2}\dot{z}\right)\dot{\rho} = 0,\label{geoz}
\eqn

\bq
\ddot{\phi} + \frac{B'}{B}\dot{\rho}\dot{\phi} = 0,\label{geophi}
\eq
where the dot denotes differentiation with respect to $\tau$.

Here we can highlight some similarities with the geodesics in the Lewis spacetime. For example, if we restrict the motion of a test particle to a fixed radius, $\rho_0$, $\dot \rho=0$, $\ddot\rho=0$, then, (\ref{geot}), (\ref{geoz}) and (\ref{geophi}) give $\ddot t =0$, $\ddot z =0$ and $\ddot \phi =0$, implying that it can not experiment any acceleration also in the $z$ or $\phi$ directions. Particularly (\ref{geophi}), which can be rewritten as

\begin {equation}
\ddot\phi=\frac{4\sigma(1-2\sigma)}{\rho}\dot\rho\dot\phi,
\end{equation}
says that the accelerated circular motion exists only if $\dot \rho\neq 0$, since $\sigma\neq0$ and $\sigma\neq 1/2$. This { behaviour} is similar to that observed in the Lewis spacetime, but for the $z$ direction. There, it was interpreted as a force which tends "to damp any motion along the z axis whenever the particle approaches that axis, and reverses this tendency, in the opposite case" \cite{228}. Here it represents the angular acceleration in a spiraled motion in the plane orthogonal to the axis.

\subsection{{Geodesics with $\dot \rho=0$}}

If we restrict the motion of a test particle to a fixed radius, $\rho_0$, $\dot \rho=0$, $\ddot\rho=0$ then, (\ref{geot}), (\ref{geoz}) and (\ref{geophi}) give us $\ddot t =0$, $\ddot z =0$ and $\ddot \phi =0$, as was pointed out before, implying that it can not experiment any acceleration also in the $z$ or $\phi$ directions, which are results similar as that found for the Lewis spacetime. The remaining equation, (\ref{georho}), gives

\begin{equation} 
\label{georhofix}
{\omega}^2-\left[\frac{1}{B^\prime}\left(A^\prime-C^\prime {v_z}^2-2k^\prime v_z\right)\right]_{\rho=\rho_0}=0,
\end{equation}
where $\omega={\phi^*}=\dot\phi/\dot t$ and $v_z=\dot z/\dot t$, for a fixed $\rho=\rho_0$, and the symbol $*$ means differentiation with respect to $t$. 

\bigskip
\subsubsection{Circular geodesics}
\bigskip

For circular geodesics we have the additional condition

\bq
\dot z =0 \Rightarrow v_z = 0.
\eq

Then, (\ref{georhofix}) gives

\bqn
\label{omega2}
{\omega}^2 &=&\left(\frac{A^\prime}{B^\prime}\right)_{\rho=\rho_0}\nonumber\\
&=&-\frac{2\rho_0^{-(n^2-3)/2}}{a n^2}\left(\frac{a^2 n^2}{n+1}\rho_0^{-n} + \frac{c^2}{n-1} \rho_0^n \right) \nb\\
&=&\frac{1}{a}\left[\frac{{\rho_0}^{8\sigma (1-\sigma)} a^2}{2 \sigma-1}+\frac{{\rho_0}^{2(1-4 \sigma^2)} c^2}{2(1-4 \sigma)^2 \sigma}\right],
\eqn

We can notice that this equation is independent of the parameter $b$, at variance with what we have in the Lewis spacetime. Moreover, we must have $\sigma \neq 0$ and $\sigma \neq \frac{1}{4}$ and $\sigma \neq \frac{1}{2}$ in order to avoid the divergence of ${\omega}^2$. 

\begin{figure}
\centering
\includegraphics[width=8cm]{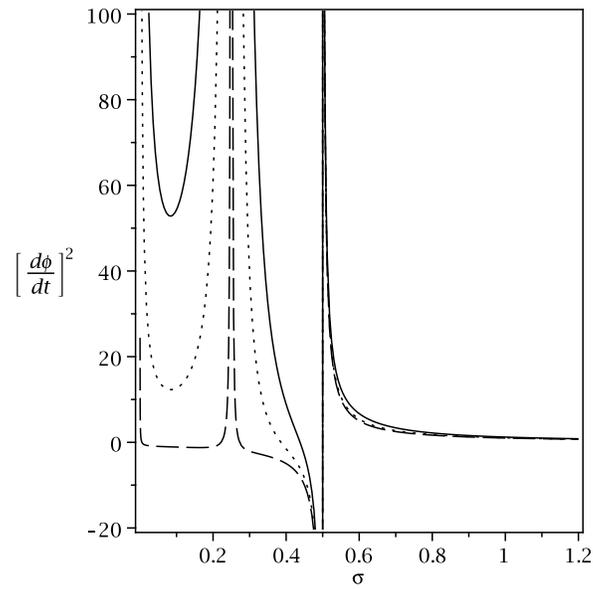}
\caption
{Plot of $\omega^{2} \equiv \left(\frac{d\phi}{dt}\right)^2$ assuming that $\rho_0=1$, $c=0.1$ (dashed line), $c=1$ (dotted line), $c=2$ (solid line), $a=1$.}
\label{figcirc0}
\end{figure}

\begin{figure}
\centering
\includegraphics[width=8cm]{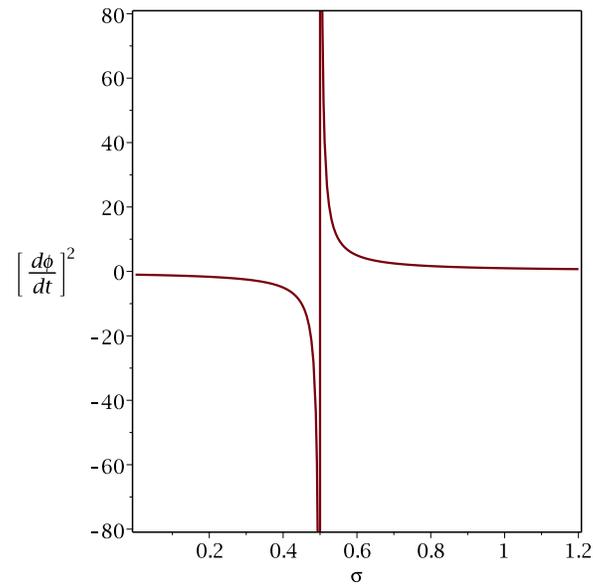}
\caption
{Plot of $\omega^{2} \equiv \left(\frac{d\phi}{dt}\right)^2$ assuming that $\rho_0=1$, $c=0, a=1$.}
\label{figcirc1}
\end{figure}

Assuming $c=0$, we get
\bq
\label{CircGeoLC}
{\omega}^2= \frac{a}{2\sigma-1} {\rho_0}^{8(1-\sigma)\sigma},
\eq
and we see, therefore, that the expression for the circular geodesic above coincides with the corresponding one for the Levi-Civita metric ($z$ and $\phi$ interchanged and $\sigma>1/2$), noting that in the present work "$a$" is equivalent to "$a^2$" and $\rho=(r\Sigma)^{1/\Sigma}$ in reference \cite{242}. Besides, comparing (\ref{omega2}) with (58) in \cite{228}, the circular geodesic in Lewis spacetime for $b=0$, we note that there the additional term introduced by the rotation of the spacetime is independent of $\rho$, while here the equivalent term depends on $\rho_0$.  Figures \ref{figcirc0} and \ref{figcirc1} show that for $c=0$ circular geodesics exist only for the interval $\sigma>1/2$, similarly to Levi-Civita geodesics (when $z$ and $\phi$ are interchanged), while for $|c|\neq 0$ there are intervals where it is possible to have circular geodesics even for $0<\sigma<1/2$.  The bigger $c$, the larger the range of allowed $\sigma$. Therefore, a non-zero $c$ parameter implies, regardless of $b$, well-defined circular geodesics for the $\sigma < 1/2$ range.

Assuming ${\omega}=0$ and $c\neq 0$,  we can check if there is any condition which allows the existence of a resting test particle. In this case we have
\bq
\label{rho0}
\rho_{\it rest} = \left[\frac{n^2 a^2 (1-n)}{c^2 (n+1)}\right]^{1/(2n)}=
\left[\frac{2(1-4\sigma)^2 a^2\sigma}{c^2(1-2\sigma)}\right]^{1/[2(1-4\sigma)]}.
\eq

This equation shows, since $\sigma \neq 1/2$, that a test particle can be at rest at some finite radius $\rho_{\it rest}$. Here we can suggest that there is an intrinsic motion of spacetime which drags the particle along in such a way that it appears to be at rest. Given a combination of the parameters $a$, $\sigma$ and $c$, only one $\rho = \rho_0$ would allow such a situation.

If $c=0$, see (\ref{CircGeoLC}), the test particle will be at rest only at the axis, for $1/2<\sigma<1$, and at an infinite radius, for $\sigma>1$.

\subsubsection{$z$-direction geodesics}
\bigskip

Now we restrict the motions to the $z$ direction. Then on the relation (\ref{georhofix}) we must impose $\dot \phi=0 \Rightarrow \omega=0$, which gives

\begin{equation}
\label{VZ}
v_z=\left(\frac{-k^{\prime}\pm\sqrt{{k^{\prime}}^2+A^{\prime}C^{\prime}}}{C^{\prime}}\right)_{\rho=\rho_0}.
\end{equation}

The expression above is very extensive in terms of the four parameter. We will therefore explore its characteristics in some important limits. Thus, similarly to what was done for the circular geodesics, we can ask here if it is possible to find a test particle at rest at any fixed $\rho=\rho_{\it rest}$ and, as expected, we find the same answer that is 

\begin{equation}
\rho_{\it rest} = \left[\frac{n^2 a^2 (1-n)}{c^2 (n+1)}\right]^{1/(2n)}=\left[\frac{2(1-4\sigma)^2 a^2\sigma}{c^2(1-2\sigma)}\right]^{1/[2(1-4\sigma)]},\nonumber
\end{equation}

Note that even in the static limit (coming back to (\ref{VZ}) and putting $b=0$ and $c=0$), that is,

\begin{equation}
\label{VzLC}
v_z=\pm \frac{a}{{\rho_0}^{n}}\,\sqrt{\frac{1-n}{1+n}}=\pm a{\rho_0}^{(4\sigma-1)}\sqrt{\frac{2\sigma}{1-2\sigma}},
\end{equation}
we can find a test particle moving in the $z$-direction with a fixed $\rho=\rho_0$, which would be a very disconcerting result. But note that it might be possible only for $\sigma<1/2$, 
 interval that is prohibited for the corresponding Levi-Civita solution ($z$ and $\phi$ interchanged).

Now considering $c=0$ and $b\neq 0$, we have

\begin{eqnarray}
v_z &=& \frac{a\left[ab(1-n)\pm{\rho_0}^n\sqrt{1-n^2}\right]}{a^2b^2(n-1)+{\rho_0}^{2n}(n+1)}\nonumber\\
&=& \frac{a\left[2ab\sigma\pm{\rho_0}^{(1-4\sigma)}\sqrt{2\sigma(1-2\sigma)}\right]}{{\rho_0}^{2(1-4\sigma)}(1-2\sigma)-2a^2b^2\sigma},
\end{eqnarray} 
which gives a complex $v_z$ for any $\sigma>1/2$, but a real value if $0\leq\sigma <1/2$.

On the other hand, for $b=0$ and $c\neq 0$, we have

\begin{eqnarray}
v_z &=&\pm \frac{c}{n} \pm \frac{a\sqrt{1-n^2}}{(1+n){\rho_0}^n}\nonumber\\
&=&\pm \frac{c}{(1-4\sigma)}\pm \frac{a\sqrt{2\sigma(1-2\sigma)}}{{\rho_0}^{(1-4\sigma)}(1-2\sigma)}, 	
\end{eqnarray}
and, as in the previous case, we have a complex $v_z$ for any $\sigma>1/2$, but a real value if $0\leq\sigma <1/2$. Similarly to what has been observed in the case of circular geodesics, we see that the presence of the $c$ parameter, and now also of the $b$ parameter, allows the existence of well-defined geodesics in the $z$ direction for models with $\sigma <1/2 $. Again, we can propose the dragging hypothesis to justify this { behaviour} of the test particle.

\subsection{{Radial geodesic}}

As another example of the new physical properties of the solution presented here, we will now consider the radial motion of a test particle in the corresponding spacetime, that is, a radial motion in a plane orthogonal to the symmetry axis implying $\dot{\phi}=\ddot{\phi}=\dot{z}=\ddot{z}=0$. Inserted into the geodesic equation (\ref{geoz}), this gives
\bqn
\frac{Ak'-A'k}{D^2}\dot{t}\dot{\rho}=\pm \frac{2c}{\rho}\dot{t}\dot{\rho}=0.
\eqn

The above equation possesses three solutions:
\begin{itemize}
\item $Ak'-A'k=0$, hence, $c=0$, without any restriction on the radial movement of the test particle. Since the geodesic equations, when $c=0$ and $\dot z =\dot\phi=0$, do not depend on the parameter $b$, it is reasonable for a particle to follow a radial trajectory, as in the static limit of Levi-Civita.

\item $\dot{\rho}=0$, which implies  $\rho=constant=\rho_0=\rho_{\it rest}$ and from (\ref{georho}), 
\begin{equation}
\rho_0=\rho_{\it rest} = \left[\frac{n^2 a^2 (1-n)}{c^2 (n+1)}\right]^{1/(2 n)}=\left[\frac{2(1-4\sigma)^2 a^2\sigma}{c^2(1-2\sigma)}\right]^{1/[2(1-4\sigma)]}.
\end{equation}
This result reinforces the possibility that a test particle can remain at rest, at the same fixed radius $\rho_0 $ as that exhibited in (\ref{rho0}). Note that it is a non-zero $c$ constant that prevents any motion of the test particle in the radial direction, either towards or away from the axis. Therefore, if there is a dragging effect on the particle, it must be closely associated with this parameter.

\item  $\dot{t}=0$, hence $t= constant$, preventing any motion of the test particle.
\end{itemize}

In the study conducted in this section we identified the existence of geodesics along the $z$-direction, as well as resting geodesics, and we can suppose that a frame dragging could occur here as was previewed by Griffiths and Santos \cite{225}. This hypothesis, which we propose, will have to be verified in the future through the junction of the vacuum, here considered, with a cylindrical source in translation.

\section{Translating cylindrical dust solution}

Here we are interested in verifying whether the spacetime described by metric (\ref{new-metric}) admits a dust solution or not. This issue is very important as a mean to distinguish between rotating  or translating spacetimes, since only in the former we expect the presence of (centrifugal) forces able to balance matter contraction, allowing the existence of such solutions. The solution for a rigidly rotating dust was obtained by van Stockum \cite{VS}. From now on, as a suitable tool, we consider metric (\ref{new-metric}) written in the form used by Griffiths and Santos \cite{225}. It is easy to see that they are equivalent. Then,

\begin{equation}
\label{A1}
ds^{2}=-A(dt-hdz)^2+d\rho^2+Cdz^2+D\rho^2d\phi^2,
\end{equation}
and the field equations for dust $\mu$ are

\begin{equation}
\label{A2}
\frac{A^{\prime\prime}}{A}-\frac{A^\prime}{2A}\left(\frac{A^\prime}{A}-\frac{2}{\rho}-\frac{C^\prime}{C}-\frac{D^\prime}{D}\right)+\frac{A}{C}{h^\prime}^2=\kappa\mu(1+2ACv^2),
\end{equation}

\begin{equation}
\label{A3}
h^{\prime\prime}+\frac{h^\prime}{2}\left(3\frac{A^\prime}{A}+\frac{2}{\rho}-\frac{C^\prime}{C}+\frac{D^\prime}{D}\right)=2\kappa\mu Cv\sqrt{1+ACv^2},
\end{equation}

\begin{equation}
\label{A4}
\frac{C^{\prime\prime}}{C}+\frac{C^\prime}{2C}\left(\frac{A^\prime}{A}+\frac{2}{\rho}-\frac{C^\prime}{C}+\frac{D^\prime}{D}\right)-\frac{A}{C}{h^\prime}^2=-\kappa\mu(1+2ACv^2),
\end{equation}

\begin{equation}
\label{A5}
\frac{D^{\prime\prime}}{D}+\frac{D^\prime}{2D}\left(\frac{A^\prime}{A}+\frac{4}{\rho}+\frac{C^\prime}{C}-\frac{D^\prime}{D}\right)+\frac{1}{\rho}\left(\frac{A^\prime}{A}+\frac{C^\prime}{C}\right)=-\kappa\mu,
\end{equation}

\begin{equation}
\label{A6}
\frac{A^\prime C^\prime}{AC}+\left(\frac{A^\prime}{A}+\frac{C^\prime}{C}\right)\left(\frac{D^\prime}{D}+\frac{2}{\rho}\right)+\frac{A}{C}{h^\prime}^2=0.
\end{equation}

Adding (\ref{A2}) and (\ref{A4}) we obtain

\begin{eqnarray}
\label{A7}
&&\frac{A^{\prime\prime}}{A}+\frac{C^{\prime\prime}}{C}+\frac{1}{\rho}\left(\frac{A^\prime}{A}+\frac{C^\prime}{C}\right)+\frac{A^\prime C^\prime}{AC}+\frac{1}{2}\frac{D^\prime}{D}\left(\frac{A^\prime}{A}+\frac{C^\prime}{C}\right)\nonumber\\
& &-\frac{1}{2}\left[\left(\frac{A^\prime}{A}\right)^2+\left(\frac{C^\prime}{C}\right)^2\right]=0,
\end{eqnarray}
which after integration produces 

\begin{equation}
(AC)^{\prime 2}D\rho^2=c_1 AC,
\end{equation}
where the integration constant $c_1=0$ because of regularity conditions along the axis $\rho=0$. Hence $(AC)^{\prime}=0$ and $AC=1$ as a consequence of the same regularity conditions. Substituting these expressions into (\ref{A6}) one obtains

\begin{equation}
\label{A8}
h=1-\frac{1}{A}.
\end{equation}

Now substituting (\ref{A7}) and (\ref{A8}) into (\ref{A2}) and (\ref{A3}) we see that there is no $v$ that satisfy both equations. Hence there is no cylindrical translating dust solution in General Relativity.

\section{Conclusions}

This paper {studies some physical and geometrical properties of} the vacuum solution (\ref{new-metric}) for a cylindrical source {supposedly in translation} along its axis of symmetry. Mathematically, this solution is akin to the Lewis solution for a cylindrically symmetric vacuum spacetime with exchanged $z$ and $\phi$ coordinates. However, we have shown that {they are physically  different}.

If the source has rigid translation, with vanishing shear, it reduces to the static Levi-Civita spacetime \cite{225}. In our case $1/2<\sigma<\infty$, and thus, $1/\sigma$, and not $\sigma$ as in the Lewis solution, is the Newtonian mass per unit length.

However, if the source is non rigidly translating, with non vanishing shear, then the integration constants do not vanish in general with $b$ and $c$ describing the non rigidity and the topological defects due to the non rigid cylindrically translating source.

As regards test particle motion in our case, well-defined circular, $z$-direction and radial geodesics exist for $\sigma$ ranges closely related to the value of the $c$-parameter, possibly including $0 < \sigma < 1/2$. Moreover, we have shown that, for a given combination of non-zero $a$, $c$ and $\sigma$ parameters, a test particle can appear at rest with respect to the coordinate 
frame here defined. We have suggested that it might be the intrinsic motion of spacetime which drags the particle along in such a way that it appears to be thus at rest.

Finally, we have shown  that no translating cylindrical dust solution can exist in the framework of General Relativity{, reinforcing our hypothesis of translation}.

Sources producing the vacuum field (\ref{new-metric}) will be studied elsewhere.

\section*{Acknowledgements}
Financial assistance from FAPERJ/UERJ (MFAdaS) is gratefully acknowledged.
The author (RC) acknowledges financial support from FAPERJ
(no. E-26/171.754/2000, E-26/171.533/2002, E-26/170.951/2006, E-26/110.432/2009
and E26/111.714/2010). The authors (RC and MFAdaS) also acknowledge
financial support from Conselho Nacional de Desenvolvimento Cient\'ifico e
Tecnol\'ogico - CNPq - Brazil (no. 450572/2009-9, 301973/2009-1 and
477268\-/2010-2). The author (MFAdaS) also acknowledges financial support
from Financiadora de Estudos e Projetos - FINEP - Brazil (Ref. 2399/03).
We also would like to acknowledge fruitful discussions with Dr. F.C. Mena.


\section*{References}

\begin{thebibliography}{100}

\bibitem{100} Brito, I., da Silva, M. F. A., Mena, F. C., and Santos, N. O. 2013 Gen. Rel. Grav. 45, 519
\bibitem {101} Levi-Civita, T. 1919 Rend. Acc. Lincei 28, 101
\bibitem {102}  Bonnor, W. B. 1992 Gen. Rel. Grav. 24, 551
\bibitem {103}  Bonnor, W. B., Griffiths, J. B., and MacCallum, M. A. H. 1994 Gen. Rel. Grav. 26, 687
\bibitem {104} Wang, A., da Silva, M. F. A., and Santos, N. O. 1997 Class. Quantum Grav. 14, 2417
\bibitem {106} da Silva, M. F. A., Herrera, L., Paiva, F. M., Santos, N. O. 1995 J. Math. Phys. 36, 3625
\bibitem {118} Dowker, J. S. 1967 Il Nuovo Cimento B52, 129
\bibitem {119} Ford, L. H. and Vilenkin, A. 1981 J. Phys. A: Math. Gen. 14, 2353
\bibitem {105} Bezerra, V. B. 1990 Annals of Phys. 203, 392
\bibitem {132} Ho, V. B., and Morgan, J. 1994 Aust. J. Phys. 47, 245
\bibitem {149} Stachel, J. 1982 Phys. Rev. D 26, 1281
\bibitem{Muriano} Muriano, A. G. R. and da Silva, M. F. A. 1997 American Journal of Physics, 65, 914
\bibitem {LewisWeyl} da Silva, M. F. A., Herrera, L., Paiva, F. M. and Santos, N. O. 1995 Gen. Rel. Grav. 27, 859
\bibitem{507} Gautreau, R. and Hoffman, R. B. 1969 Nuovo Cimento B 61, 411.
\bibitem {160} Linet, B. 1986 J. Math. Phys. 27,  1817
\bibitem {170} Tian, Q. 1986 Phys. Rev. D 33, 3549
\bibitem {180} da Silva, M. F. A., Wang A., Paiva, F. M. and Santos N. O. 2000 Phys. Rev. D 61, 044003
\bibitem {190} Griffiths, J., Podolsky, J. 2010 Phys. Rev. D 81, 064015
\bibitem {200} Bezerra de Mello, E. R., Brihaye, Y. and Hartmann, B. 2003 Phys. Rev. D 67, 124008
\bibitem {221} Bhattacharya, S. and Lahiri, A. 2008 Phys.Rev. D 78, 065028
\bibitem {500} Brito, I., da Silva, M. F. A., Mena, F. C. and Santos, N. O. 2015 Class. Quantum Grav. 32, 185015
\bibitem {31} Lewis, T. 1932 Proc. R. Soc. London 136, 176
\bibitem {33} Levi-Civita, T. 1917 Rend. Acc. Lincei, 26, 307
\bibitem {34} da Silva, M. F. A., Herrera, L., Paiva, F. M. and Santos, N. O. 1995 Gen. Rel. Grav. 27, 859
\bibitem {35} Stachel, J. 1982 Phys. Rev. D 26, 1281 Frehland, E. 1971 Comm. Math. Phys. 23, 127 Bonnor, W. B. 1980 J. Phys.A 13, 2121
\bibitem {36} Herrera, L., Paiva, F. M. and Santos, N. O. 2000 Class. Quantum Grav. 17, 1549
\bibitem{Krasinski} Krasinski, A. 1975 Acta Phys. Polon. B 6, 223
\bibitem{Santos} Santos, N. O. 1993 Class. Quantum Grav. 10, 2401; 1997 Class. Quantum Grav. 14, 3177
\bibitem{Einstein} Einstein, A., and Rosen, N. 1937 J. Franklin Inst. 223, 43
\bibitem {229} Herrera, L., and Santos, N. O. 2005 Class. Quantum Grav. 22, 2407 Herrera, L., MacCallum, M. A. H. and Santos, N. O. 2007 Class. Quantum Grav. 24, 1033
\bibitem {217} Di Prisco, A., Herrera, L., MacCallum, M. A. H. and Santos, N. O. 2009 Phys. Rev. D 80, 064031
\bibitem{Herrera1} Herrera, L., Paiva, F. M., and Santos, N. O. 1999 J. Math. Phys. 40, 4064
\bibitem{Herrera2} Herrera, L., Paiva, F. M., and Santos, N. O. 2000 Int. J. Mod. Phys. D 9, 649
\bibitem {225} Griffiths, J. B., and Santos, N. O. 2010 Int. J. Mod. Phys. D 19, 79
\bibitem {Godel} Godel, K. 1949 Rev. Mod. Phys. 21, 447
\bibitem {BSM} Bonnor, W. B, Santos, N. O.  and MacCallum, M. A. H. 1998 Class. Quantum Grav. 15, 357
\bibitem {GS10} Griffiths, J. B. and Santos, N. O. 2010 Class. Quantum Grav. 27, 125004
\bibitem {228} Herrera, L., and Santos, N. O. 1998 J. Math. Phys. 39, 3817
\bibitem {242} Herrera, L., Santos, N. O., Teixeira, A. F. F. and Wang, A. Z., Class.Quant.Grav. 18 (2001) 3847
\bibitem {VS} van Stockum, W. J., Proc. R. Soc. Edin. 57, (1937) 135
{
\bibitem{Bronnikov2019} Bronnikov, K., Santos, N.O. and Wang, A. 2019  	[arXiv:1901.06561]
}

\end{thebibliography}
\end{document}